# Resistive switching in MoSe$_2$/BaTiO$_3$ hybrid structures

J. P. B. Silva,[1,2,*] C. Almeida Marques,[3,4] J. Agostinho Moreira,[2] O. Conde[3,4,*]

[1]Centre of Physics, University of Minho, Campus de Gualtar, 4710-057 Braga, Portugal

[2]IFIMUP and IN-Institute of Nanoscience and Nanotechnology, Departamento de Física e Astronomia, Faculdade de Ciências da Universidade do Porto, 4169-007 Porto, Portugal

[3]Departamento de Física, Faculdade de Ciências, Universidade de Lisboa, 1749-016 Lisboa, Portugal

[4]CeFEMA – Center of Physics and Engineering of Advanced Materials, Universidade de Lisboa, 1749-016 Lisboa, Portugal

* Authors to whom correspondence should be addressed. Electronic addresses: josesilva@fisica.uminho.pt and omconde@ciencias.ulisboa.pt

Here we study the resistive switching (RS) effect that emerges when ferroelectric BaTiO$_3$ (BTO) and few-layers MoSe$_2$ are combined in one single structure. The C-V loops reveal the ferroelectric nature of both Al/Si/SiO$_x$/BTO/Au and Al/Si/SiO$_x$/MoSe$_2$/BTO/Au structures and the high quality of the SiO$_x$/MoSe$_2$ interface in the Al/Si/SiOx/MoSe$_2$/Au structure. Al/Si/SiO$_x$/MoSe$_2$/BTO/Au hybrid structures show the electroforming free resistive switching that is explained on the basis of the modulation of the potential distribution at the MoSe$_2$/BTO interface via ferroelectric polarization flipping. This structure shows promising resistive switching characteristics with switching ratio of $\approx 10^2$ and a stable memory window, which are highly required for memory applications.

The emergency of two dimensional (2D) materials, such as graphene and transition metal dichalcogenides (TMDs), has paved the way to the recent field of nano-scale electronics.[1,2] In particular, semiconductor TMDs with chemical formula $MX_2$ (with M=Mo,W and X=S, Se) have attracted much attention due to their tunable electronic properties, dependent upon the number of layers in the material. As the number of layers decreases, the band structure of the TMD changes so that the indirect band gap in bulk and few-layer TMD crystals transforms into a direct band gap in the monolayer.[3] Therefore, devices such as field-effect transistors, photo-detectors, chemical sensors, and solar cells, have been successfully fabricated using a TMD based device structure.[4-7]

Recently, new devices combining TMDs with ferroelectrics have emerged as a new hotspot for promising applications in electronics and optoelectronics. Most of these studies focus on the memory and logic applications of ferroelectric field-effect transistors with TMDs channels, the ferroelectric material acting as gate dielectrics.[8-11] However, charge carrier modulation within the TMDs due to ferroelectric polarization of the ferroelectrics has been insufficiently investigated.

On another hand, no studies regarding the combination of ferroelectrics and TMDs for resistive random access memories (RRAMs) have been published so far. In this letter, we study the resistive switching (RS) effect in TMDs/ferroelectric hybrid structures. We have chosen barium titanate ($BaTiO_3$ – BTO) as the ferroelectric material due to its eco-friendly nature, excellent spontaneous polarization, as large as 26 μC/cm$^2$, and moderate coercive field (≈1 kV/cm).[12] From all the TMDs, most studies have focused on $MoS_2$. However, other TMDs such as $MoSe_2$ possess attractive properties. For instance, monolayer $MoSe_2$ has a direct bandgap of 1.5 eV, which is close to the optimal bandgap value of single-junction solar cells and photoelectrochemical cells.[13]

Moreover, few-layer MoSe$_2$ has nearly degenerate indirect and direct bandgaps, and an increase in temperature can effectively push the system towards the 2D limit.[14]

Yet, there are no reports on the RS characteristics of MoSe$_2$/BTO hybrid structures, neither on the polarization coupling at MoSe$_2$/BTO heterojunctions.

Here we report on the RS effect in MoSe$_2$/BTO hybrid structures deposited on Si/SiO$_x$ substrates. The relation between the ferroelectric polarization and the RS effect was investigated and the mechanism underlying this effect is highlighted. In order to assess the performance of Al/Si/SiO$_x$/MoSe$_2$/BTO/Au structures for non-volatile memory applications, endurance tests were also carried out.

Three types of samples, namely BTO, MoSe$_2$ and MoSe$_2$/BTO hybrid structures, were grown on top of Si/SiO$_x$ substrates. The 87 nm thick SiO$_x$ dielectric layers were prepared by the dry oxidation of *p*-doped (111) silicon wafer pieces (resistivity $< 5 \times 10^{-3}$ Ωcm). MoSe$_2$ nanolayers were grown onto the Si/SiO$_x$ substrates by chemical vapor deposition (CVD), using a similar procedure as reported by Shaw *et al.*.[15] MoO$_3$ and Se powders were heated up to 800 ºC and 300 ºC, respectively, and the growth temperature was 790 ºC. The background gas consisted of Ar + H$_2$ flowing along the tube furnace at a rate of 60 sccm.[16] BTO films, with thickness of 170 nm, were grown by ion beam assisted sputter deposition (IBSD), using a commercially available BaTiO$_3$ target.[12] During the deposition, the substrate was kept at a temperature of 330 ºC. Subsequently, the samples were annealed in vacuum (≈2.0×10$^{-5}$ mbar) at 650 ºC for 30 minutes in order to improve their crystallinity.[17]

The surface morphology of the samples was analyzed by field emission scanning electron microscopy (FE-SEM) and atomic force microscopy (AFM) in tapping mode. The height profile presented is the average curve of five different measurements performed along equivalent paths in the AFM image. The few-layers MoSe$_2$ were structurally characterized by Raman microprobe spectrometry, with 532 nm excitation source, while X-ray diffraction (XRD) in $\theta$–$2\theta$ coupled mode, using CuKα radiation, was performed for all single layers and hybrid structures. For the electrical characterization of the samples, top circular gold (Au) electrodes, with a diameter of 1 mm, were deposited by thermal evaporation on the upper surface, while bottom Al-electrodes were attached to the Si wafer backside by electric spark. The capacitance–voltage (C–V) characteristics were measured using a precision LCR meter at an ac voltage of 50 mV and frequency of 10 kHz. Current-voltage (I-V) characteristics were measured using a programmable electrometer.

Figure 1(a) shows the FE-SEM surface microstructure of the Si/SiO$_x$/MoSe$_2$ layer revealing the presence of triangles of lateral size mainly in the range 750 – 800 nm, whereas few smaller triangles of about 450 – 500 nm size can also be seen. This triangular morphology is typical of mono- and few-layers MoSe$_2$.[13] Thus, to get an insight on the number of layers, AFM images and micro-Raman spectra were recorded in the same region as the SEM image. The height profile seen on the inset of the AFM image (Fig. 1(b)) was taken along the green line crossing the edge of one of the triangles, yielding a height of approximately 1.8 nm, which corresponds to ≈ 2-layers MoSe$_2$. This result is corroborated by the micro-Raman spectrum (Fig. 1(c)) acquired over one of the larger triangles of Fig. 1(a). Comparing to the Raman spectrum of the substrate, the Si/SiO$_x$/MoSe$_2$ structure has additional peaks at 170.3, 241.1, 286.1 and 352.1 cm$^{-1}$, which can be assigned to the phonon modes of low dimensional MoSe$_2$, respectively, $E_{1g}$, $A_{1g}$, $E_{2g}^1$, $B_{2g}^1$.[18,19] The frequency difference between the $E_{2g}^1$ and $A_{1g}$ modes is 45.0 cm$^{-1}$, which is close to the reported value for MoSe$_2$ monolayer.[18] This result along with the weak intensity of the $B_{2g}^1$ mode suggests that these isolated large triangles are few-layers of MoSe$_2$.[13,20] Moreover, low intensity additional peaks marked with "*" at about 147, 257, 303, 432, and 451 cm$^{-1}$ can be seen in the spectrum, which might be associated to Se[21] and Mo$_4$O$_{11}$.[22]

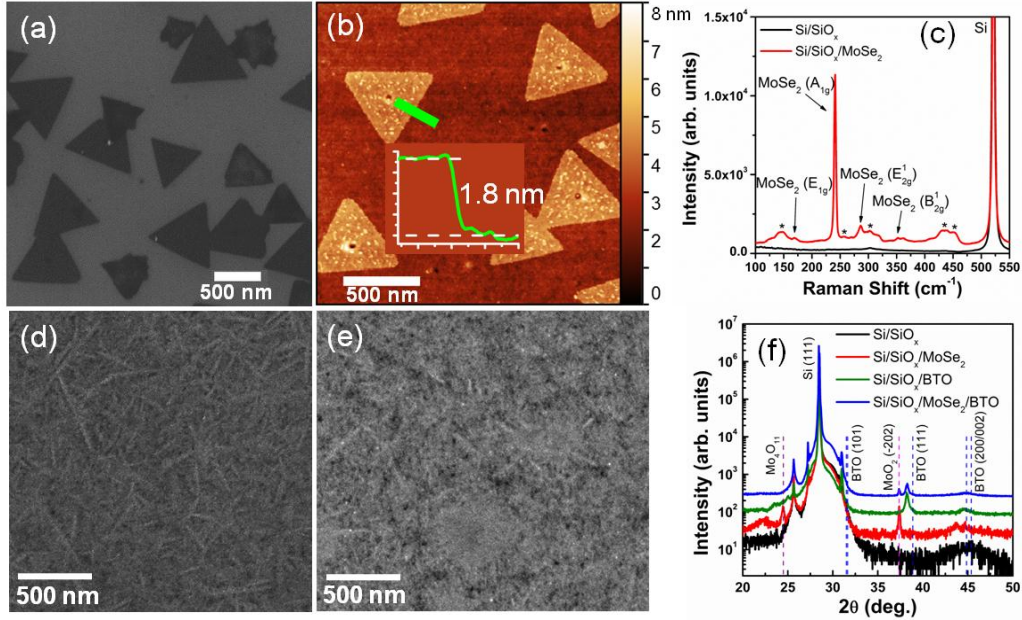

FIG. 1. (a) FE-SEM micrograph of Si/SiO$_x$/MoSe$_2$ single layer; (b) AFM image of the sample in (a) with the inset showing the height profile across the edge of one triangle (green line); (c) micro-Raman spectra recorded with the laser beam focused onto one of the larger triangles (red curve) and of the oxidized substrate (black curve); FE-SEM micrographs of (d) Si/SiO$_x$/BTO and (e) Si/SiO$_x$/MoSe$_2$/BTO structures; (f) XRD patterns of the three structures and the oxidized substrate.

Figures 1(d) and (e) show the FE-SEM micrographs of the BTO terminated structures, respectively Si/SiO$_x$/BTO and Si/SiOx/MoSe$_2$/BTO, which reveal different BTO film morphologies depending on the composition of the samples. While Fig. 1(d) displays dense whisker-type morphology, with the elongated shapes randomly oriented, in Fig. 1(e) the same microstructure appears combined with another type consisting of variable sized clusters. It is apparent that whiskers grow on the surface regions between the MoSe$_2$ triangles while the clusters nucleate mainly on the underneath triangles.

In Fig. 1(f) we have plotted the XRD patterns of the three different structures studied in this letter, as well as, of the oxidized substrate for comparison. The XRD patterns of the Si/SiO$_x$/BTO and Si/SiO$_x$/MoSe$_2$/BTO structures exhibit the characteristic (101) and (111) Bragg peaks of the tetragonal perovskite phase of BTO, marked with dashed blue lines, without any secondary phase.[23] In addition, compared to the position of the Bragg peaks of the JCPDS card no. 5-0626, the reflections of BTO in both structures are shifted to lower $2\theta$ values due to the tensile stress caused by the thermal expansion mismatch with the Si substrate.[24] Also, a weak peak assigned to monoclinic $MoO_2$ is found at $\approx 37.3°$ in both structures containing the $MoSe_2$, while a peak corresponding to $Mo_4O_{11}$, at $2\theta \approx 24.5°$, is visible only in the $Si/SiO_x/MoSe_2$ structure. The latter corroborates the Raman assignment above mentioned. The XRD results indicate that there are residual molybdenum oxide intermediate phases in the samples, which suggest that there are still reactions to occur. Within the resolution of the technique, the presence of 2H-$MoSe_2$ could not be detected by XRD. This was expected due to the very thin $MoSe_2$ layers, as recently reported.[25]

Figures 2(a)-(c) show the capacitance-voltage (C–V) characteristics of $Al/Si/SiO_x/MoSe_2/Au$, $Al/Si/SiO_x/BTO/Au$ and $Al/Si/SiO_x/MoSe_2/BTO/Au$ structures, respectively.

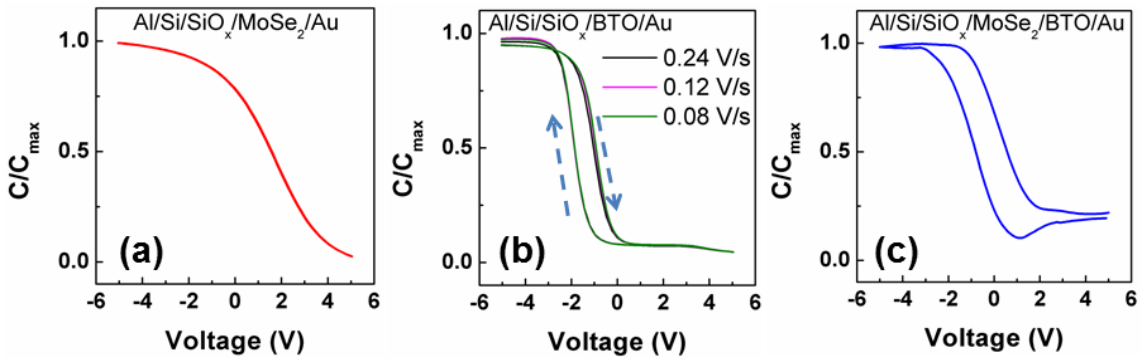

FIG. 2. C-V characteristic curves of (a) $Al/Si/SiO_x/MoSe_2/Au$, (b) $Al/Si/SiO_x/BTO/Au$ and (c) $Al/Si/SiO_x/MoSe_2/BTO/Au$ structures. Different scanning speeds of the bias voltage were used in (b).

The gate voltage was swept from -5 V to +5 V and then backs again to -5 V. Figure 2(a) shows the C-V characteristics for the $Al/Si/SiO_x/MoSe_2/Au$ structure. A transition from accumulation to depletion is clearly observed and no hysteresis can be found suggesting that the $SiO_x/MoSe_2$ interface is of good quality.[26]

The C-V plot for Al/Si/SiO$_x$/BTO/Au structure is shown in Fig. 2(b) and exhibits a clockwise hysteresis loop, as indicated by the arrows, suggesting that the memory window ($\approx$1.0 V) is induced by the ferroelectric polarization switching.[27,28] When a negative voltage is applied to the top electrode, most of the applied voltage drops on BTO and SiO$_x$ layers. The substrate contribution can be neglected and the p-Si is in the accumulation mode. When the applied voltage is decreased, the Si surface is depleted and thus, the capacitance decreases and remains constant in the strong inversion region. Nevertheless, the clockwise hysteresis might also be a result of mobile charges at the interfaces.[29] In order to confirm that this effect can be discarded, the C-V was measured by changing the scanning speed of the bias voltage. Fig. 2(b) evidences the C-V curves for sweeping speeds of 0.08V/s, 0.12V/s and 0.24 V/s. The C-V curves do not change remarkably with the different sweeping modes, which allow us to infer that the effect of ionic mobile charges is negligible. Therefore, the memory window is due to polarization flipping. Moreover, the memory window is theoretically equal to twice the coercive voltage[30] and the value found for the coercive field (Ec) is $\approx$33 kV/cm.

Usually the insertion of a 2D material, like graphene, leads to smaller remnant ferroelectric polarization, which can be explained by the lower carrier density of the 2D material being unable to compensate the polarization completely.[31] The C-V curve for the Al/Si/SiO$_x$/MoSe$_2$/BTO/Au structure also exhibits a hysteresis effect similar to the one observed for the Al/Si/SiO$_x$/BTO/Au structure. However, the memory window in this case is slightly larger and is equal to $\approx$1.4 V. It is possible that some of the ferroelectric domains are pinned in one orientation at the MoSe$_2$/BTO interface due to a strong interaction between MoSe$_2$ and the ferroelectric domains.[32] In this case, some of the ferroelectric domains cannot be easily reoriented and should result in the increase of the coercive field and consequently in the memory window. However, the memory window is still present confirming the ferroelectric nature of the BTO film.

Figures 3(a)-(c) depict the room temperature I-V characteristics of Al/Si/SiO$_x$/MoSe$_2$/Au, Al/Si/SiO$_x$/BTO/Au and Al/Si/SiO$_x$/MoSe$_2$/BTO/Au structures, wherein I stands for the modulus of the current.

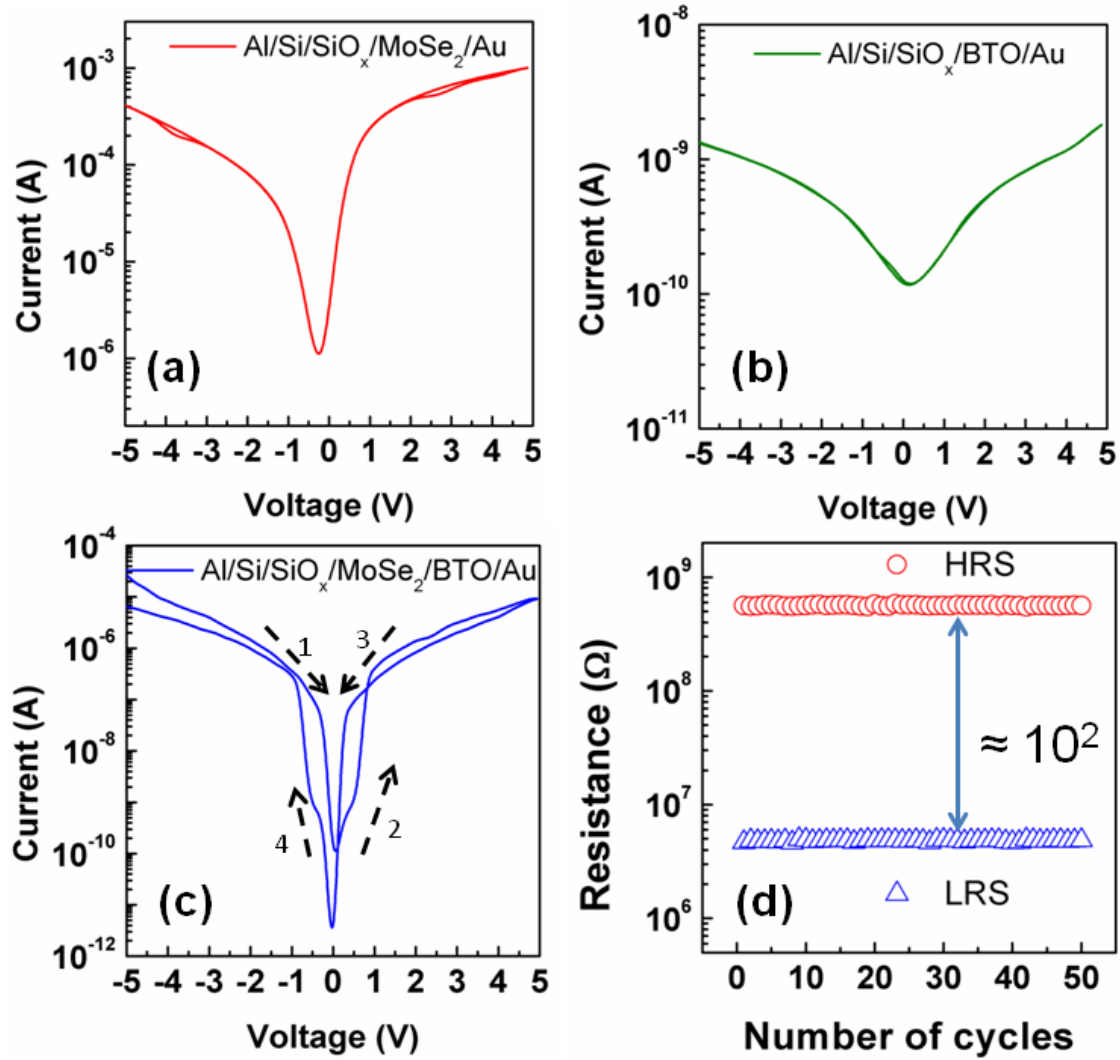

FIG. 3. I-V characteristic curves of (a) Al/Si/SiO$_x$/MoSe$_2$/Au, (b) Al/Si/SiO$_x$/BTO/Au and (c) Al/Si/SiO$_x$/MoSe$_2$/BTO/Au structures. (d) Resistance evolution for each resistance state as a function of the number of cycles.

The dc voltage V was applied on the bottom Al electrode with the top Au electrode as the ground. The dc voltage was first swept from -5 V to +5 V, and then in the reverse direction. The I-V curves of Al/Si/SiO$_x$/MoSe$_2$/Au and Al/Si/SiO$_x$/BTO/Au structures are shown in Fig. 3(a) and (b) and exhibit a diode behavior, with a rectification ratio of 2.6 and 1.5, respectively. The diode behavior suggests that a p-n heterojuntion is formed at p-Si/MoSe$_2$ and p-Si/BTO.[1] In fact, unlike graphene, TMDs, such as MoS$_2$ and MoSe$_2$, usually exhibit n-type dominant transport behavior.[1,2,9] Moreover, the leakage currents in BTO films are usually attributed to the inevitable oxygen vacancies and, thus, BTO is also assumed as n-type semiconductor.[12] However, significant difference can be observed in the I-V curves when both MoSe$_2$ and BTO films are join together in one structure. Remarkably, the hysteresis effect is much more pronounced in Al/Si/SiO$_x$/MoSe$_2$/BTO/Au structures.

As illustrated in Fig. 3(c), the I-V curve shows the bipolar RS effect, without any electroforming process. In fact, the internal electric field in each MoSe$_2$ and BTO layers of the Al/Si/SiO$_x$/MoSe$_2$/BTO/Au structure is equal or smaller when compared with the one in the individual thin layers of the Al/Si/SiO$_x$/MoSe$_2$/Au and Al/Si/SiO$_x$/BTO/Au structures. If any electroforming occurred, RS would also be visible in Al/Si/SiO$_x$/MoSe$_2$/Au and/or Al/Si/SiO$_x$/BTO/Au structures. It has been previously suggested that the switchable diode effect in ferroelectric films is caused by the polarization modulation of Schottky barriers.[33] It is observed that the switching from high resistance state (HRS) to low resistance state (LRS) occurred at threshold voltage of ±0.6 V. The polarization switching of BTO layer in hybrid structure causes the positive or negative polarization charges at the interface due to charge coupling at MoSe$_2$/BTO interface. This leads to the formation of high and low current states, where a change in the conductivity occurs when the electron carriers in the MoSe$_2$ are in accumulation or depleted at the interface.

As schematically shown in Fig. 4(a), for positive polarization charges at the interface, electron carriers in MoSe$_2$ tend to accumulate at the interface and thus the hybrid structure is highly conductive. For the negative polarization charges at the interface, electron carriers are depleted from the interface and thus the hybrid structure becomes low conductive, as shown in Fig. 4(b). The resistance of two states, such as low resistance state (LRS) and high resistance state (HRS), was read-out at -0.4 V and the RS ratio [$R_{HRS}/R_{LRS}$] was found to be ≈$10^2$, which is higher than the ones reported in literature for MoSe$_2$ based devices, where the RS is related with oxygen vacancies migrating from the substrate or formation of conductive filaments.[34,35] In the present hybrid structures, the resistive switching can be well explained based on the modulation of the potential distribution at the MoSe$_2$/BTO interface via ferroelectric polarization flipping. Moreover, the similitude of coercive field ($E_c$) = 41±5 kV/cm and switching field ($E_s$) = 35±5 kV/cm confirms the strong coupling between the resistive switching and the polarization switching.[36]

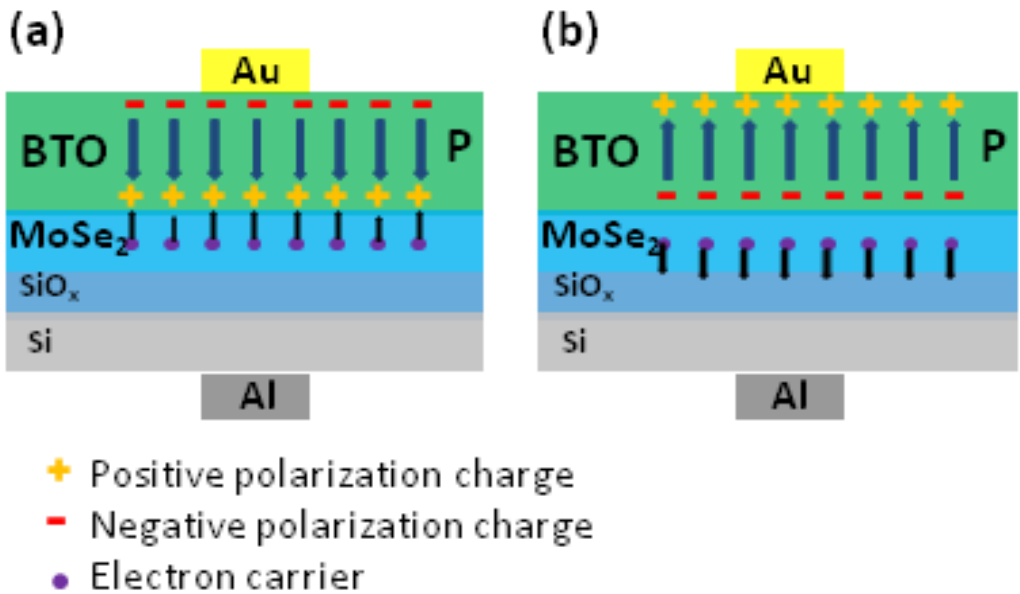

FIG. 4. Schematic charge distribution for Al/Si/SiO$_x$/MoSe$_2$/BTO/Au in **(a)** accumulation and **(b)** depletion

To evaluate the possible application in memory devices, the endurance characteristics of the Al/Si/SiO$_x$/MoSe$_2$/BTO/Au structures were investigated. The resistance evolution of HRS and LRS is shown in Fig. 3(d). It is clearly noticed that HRS and LRS are stable after 50 cycles and therefore, the hybrid structures display a reliable resistive switching behavior.

In this work, it was demonstrated the switchable diode effect in the $MoSe_2$/BTO hybrid structures. The origin of the switchable diode effect is attributed to charge coupling at $MoSe_2$/BTO interface, which might be related to the ferroelectric polarization reversal process. The coexistence of C–V hysteresis and RS characteristics and the equivalence of coercive field ($E_c$) and switching field ($E_s$) confirm the coupling between ferroelectric polarization reversal and RS effect. Our results therefore suggest that the $MoSe_2$/BTO hybrid structures are promising candidates for non-volatile ferroelectric resistive memories.

**Acknowledgments**

This work was supported by: (i) Portuguese Foundation for Science and Technology (FCT) in the framework of the Strategic Funding Contracts UID/FIS/04650/2013 and UID/CTM/04540/2013; (ii) Project Norte-070124-FEDER-000070 Nanomateriais Multifuncionais. The author J.P.B.S. is grateful for financial support through the FCT Grant SFRH/BPD/92896/2013. C.A.M. acknowledges a scholarship funded by the UID/CTM/04540. The authors would also like to thank Engineer José Santos for technical support at Thin Films Laboratory.